\documentclass[showkeys,12pt,a4paper]{article}
\title{Moduli and BPS configurations of the BLG theory}
\author{
Shankhadeep Chakrabortty
\thanks{Institute of Physics, Bhubaneswar, India.
        Email: \href{mailto:sankha@iopb.res.in}{\tt sankha@iopb.res.in}}
\and
Sudipto Paul Chowdhury
\thanks{Department of Theoretical Physics, Indian Association for the Cultivation of Science, Calcutta-700032, India.
        Email: \href{mailto:tpspc@iacs.res.in}{\tt tpspc@iacs.res.in}}
\and
Thomas K\"oppe
\thanks{School of Mathematics, University of Edinburgh, Scotland, UK.
        Email: \href{mailto:t.koeppe@ed.ac.uk}{\tt t.koeppe@ed.ac.uk}}
\and
Koushik Ray
\thanks{Department of Theoretical Physics, Indian Association for the Cultivation of Science, Calcutta-700032, India.
        Email: \href{mailto:koushik@iacs.res.in}{\tt koushik@iacs.res.in}}
}
\date{}

\usepackage{microtype,cmap,amsmath,amssymb,bm}
\usepackage[margin=3cm]{geometry}
\usepackage[dvipsnames]{xcolor}
\usepackage[colorlinks=true, linkcolor=RawSienna, urlcolor=OliveGreen, citecolor=NavyBlue,
            pdftitle={BPS configurations of the BLG theory},
            pdfauthor={Shankhadeep Chakrabarty, Sudipto Paul Chowdhury, Thomas Köppe, Koushik Ray},
            pdfsubject={Mathematical Physics, String Theory}, pdfproducer={Pure Magic}, pdfcreator={Math2PDF}]{hyperref}

\numberwithin{equation}{section}

\DeclareMathOperator{\tr}{Tr}
\DeclareMathOperator{\spec}{Spec}
\DeclareMathOperator{\spin}{Spin}
\newcommand{\ternary}[1]{{\langle\,#1\,\rangle}}
\newcommand{\viz}{\textit{viz.\ }}

\newcommand{\R}{\mathbf{R}}

\newcommand{\G}{\Gamma}
\newcommand{\rt}{\longrightarrow}
\newcommand\pa{\partial}

\begin{document}

\maketitle

\begin{abstract}
\noindent We study the moduli space of scalars in the BLG
theory with and without a constant background four-form field. 
The classical vacuum moduli space is sixteen-dimensional in the
absence of the four-form field. In its presence, however, the 
moduli space of BPS configurations may be reduced in dimension. We
exemplify this with a BPS configuration having $SO(1,2)$ world-volume
symmetry and $SO(4) \times SO(4)$ R-symmetry in the presence of a four-form
field, by constructing an explicit solution.
\end{abstract}

\clearpage

\section{Introduction}
The BLG theory, named after its inventors \cite{bagger1, bagger2,
bagger3, bagger4, gustav}, is the maximally supersymmetric
three-dimensional gauge theory with matter based on a ternary algebra.
The BLG theory contains a Chern-Simons gauge field, eight scalars and
a Majorana-Weyl spinor. The action has sixteen supersymmetries
\cite{schwarz}. The global R-symmetry of the action is $SO(8)$; the
eight scalars are thereby interpreted as the eight transverse
directions to the three dimensions of the gauge theory. The gauge
field as well as the matter fields are valued in a ternary algebra,
satisfying the so-called fundamental identity. We shall concern
ourselves with a realization of the theory furnished by a completely
antisymmetric ternary product and a Euclidean metric. The BLG theory
is deemed to describe the world-volume theory of M2-branes of the
eleven-dimensional M-theory. Among the reasons for this expectation
are its having sixteen supersymmetries, likely to be superconformal
\cite{schwarz}, and the existence of eight real scalars which may be
interpreted as the eight directions transverse to an M2-brane. A
third argument in flavor of such an interpretation comes from the
existence of a limit in which the BLG theory yields the
super-Yang-Mills theory of D2-branes in the leading order in
gauge-coupling, the latter being the vacuum expectation value of one
of the eight scalars, thereby retaining only seven transverse
directions to the D2-branes, as required \cite{mukhi1}. Various
aspects of the BLG theory as well as its variations have been studied
\cite{mukhi1, mukhi2, mukhi3, mukhi4, raam, figu2, pass, cherkis,
singh, palmkvist, sudipto,chetan}. More recently, the BLG theory has been
generalized to describe M2-branes on non-flat backgrounds containing a
constant four-form field which contributes to the action of the theory
with a mass term for the scalar fields and the fermion, as well as a
flux term \cite{lambert2}. In another line of development, BPS
configurations of the BLG theory have been studied leading to their
classification based on R-symmetry breaking \cite{Kim,gustav2},
following a similar classification in Yang-Mills theories
\cite{Bak,loginov}.

Here we first study the classical vacuum moduli space of the BLG
theory without the four-form field. The moduli space is written
algebraically in terms of gauge-invariant quantities, furnishing a
global description. Upon considering the scalar degree of freedom arising
from the three-dimensional gauge field, the dimension of the moduli space 
is sixteen. We then analyze a BPS configuration of the modified theory
with world-volume Lorentz symmetry $SO(1,2)$ and an $SO(4) \times
SO(4)$ R-symmetry in terms of the gauge-invariant variables 
to obtain an explicit solution for the scalars. 

In the next section we write down the action and supersymmetry
transformations to set up notation. In Section~\ref{vac:mod} we study
the classical vacuum moduli space using gauge-invariant variables. In
Section~\ref{so4:sol} we use the gauge-invariant variables to show
that the moduli space reduces to a point in the presence of the
four-form field. We then find an explicit configuration, before
concluding in Section~\ref{concl}.


\section{BLG theory}\label{blg:rev}
Let us begin with a discussion of some features of the BLG theory and
its deformation by a background four-form field that will be used
later. The BLG theory is an $\mathcal{N} = 8$ supersymmetric theory
in three dimensions, given by the Lagrangian
\begin{multline}\label{action1}
  \mathcal{L} = \tr \bigl(-\textstyle\frac{1}{2}(D_\mu X^{I})(D^\mu X^{I}) + 
\textstyle\frac{i}{2}\bar\Psi\gamma^\mu D_\mu \Psi
   + \textstyle\frac{i}{4}\bar\Psi \Gamma_{IJ} \ternary{X^I, X^J, \Psi} - 
\textstyle\frac{1}{12} \ternary{X^I, X^J, X^K}^2 \bigr) \\
   + \textstyle\frac{1}{2}\epsilon^{\mu\nu\lambda} \bigl( f^{abcd}A_{\mu ab}\partial_\nu A_{\lambda cd}
   + \textstyle\frac{2}{3}f^{cda}{}_gf^{efgb} A_{\mu ab}A_{\nu cd}A_{\lambda ef} \bigr). \qquad\qquad\qquad\ \quad
\end{multline}
Here $\mu = 0, 1, 2$ designates the world volume directions, $I =
1,\dotsc, 8$ indexes the flavors and $a = 1,2,3,4$ the gauge algebra.
$X^I_a$, $\Psi_a$ and $A_{\mu ab}$ are the scalars, the Majorana-Weyl
spinor and the gauge field, respectively. $\gamma$ and $\G$ are,
respectively, the three- and eight-dimensional gamma matrices. 
The structure constants of
the ternary algebra are denoted by $f^{abcd}$, while the ternary
bracket is written as $\langle~,~,~\rangle$. The repeated indices are
summed in the above expression and in the following unless stated
otherwise. Denoting the generators of the ternary algebra as
$\tau_a$, the metric tensor raising and lowering gauge indices is
written as
\begin{equation}
  h_{ab} = \tr \tau_a \tau_b.
\end{equation} 
We use the generators to write the fields valued in the ternary algebra as
\begin{gather}
\label{repscalar}
  X^I = h^{ab}X^I_a \tau_b, \\
\label{repfermion}
  \Psi = h^{ab}\Psi_a \tau _b.  
\end{gather}
The action obtained from the Lagrangian \eqref{action1} is invariant
under the supersymmetry transformations \cite{bagger3}
\begin{gather}
\label{susyscalar}
  \delta X^I = i \; \overline{\theta} \; \Gamma^I \Psi, \\
\label{susyfermion}
  \delta \Psi = D_\mu X^I \gamma_\mu \Gamma^I \theta -
  \frac{1}{6}\Gamma^{I J K} \ternary{X^I, X^J, X^K} \, \theta, \\
\label{susygauge}
  \delta A_{\mu}(\phi) = i \; \overline{\theta} \, \gamma_\mu \Gamma^I \ternary{\Psi, X^I, \phi},
\end{gather}
where $\phi$ represents either a $X^I$ or $\Psi$ and $\theta$ denotes the
parameter of supersymmetry variation.  The supersymmetry
transformations close on-shell up to translation and gauge
transformation. A realization of the BLG theory, indeed, the only
known finite-dimensional representation, is furnished by an $SO(4)$
gauge theory, in which the $X^I$ and the fermion $\Psi$ transform as
vectors under the gauge group \cite{bagger1,bagger3,papado,gutow}.  
In this instance the structure
constant is taken to be the rank-four antisymmetric tensor
\begin{equation}\label{strucconst}
  f^{a b c d} = \epsilon^{a b c d},
\end{equation} 
where we have set the level of the Chern Simons action to be unity, 
while the metric is taken to be Euclidean, 
\begin{equation}\label{symbilin}
  h_{ab} = \delta_{ab}.
\end{equation}
The ternary bracket then reads
\begin{equation}\label{algebra}
  \ternary{X^I,X^J,X^K} = \epsilon^{abcd} X^I_a X^J_b X^K_c \tau_d.
\end{equation}
Henceforth we shall only consider this realization and refer to it as
``the BLG theory".

The BLG theory has been generalized to incorporate a constant
self-dual four-form field $G$ \cite{lambert2}. The four-form field
satisfies a self-duality condition
\begin{equation}\label{self-dual}
  \widetilde{G }_{IJKL}=G_{IJKL}, 
\end{equation} 
where the dual of the four-form field $G$ is defined as
\begin{equation}
 \widetilde{G}_{I J K L} = \frac{1}{4!} \epsilon_{I J K L P Q R S} \, G^{P Q R S}.
\end{equation}
Incorporation of the four-form to the BLG theory is effected by adding a mass
deformation term and a flux term to the Lagrangian. The modified Lagrangian
reads \cite{lambert2}
\begin{equation}\label{action2}
  \mathcal{L}' = \mathcal{L} + \mathcal{L}_\text{mass} +
  \mathcal{L}_\text{flux},
\end{equation} 
where the extra terms are
\begin{gather}
\label{massterm}
  \mathcal{L}_\text{mass} = -\frac{1}{2}m^2 \delta^{IJ} \tr(X^I X^J) 
  + \tr(\overline{\Psi} \,  \Gamma^{I J K L} \, \Psi) \widetilde{G}_{I J K L}, \\
\label{fluxterm}
  \mathcal{L}_\text{flux} = - c \; \widetilde{G}_{I J K L} \tr(X^I \ternary{X^J, X^K, X^L}),
\end{gather}
where the mass $m$ is determined by the four-form field as
$m^2 = \frac{c^2}{768} G^2$, with $G^2 = G^{IJKL} G_{IJKL}$, while
$c$ is an arbitrary parameter.

The four-form field $\widetilde{G}_{IJKL}$ contributes to the
supersymmetry transformation of the fermion with a term linear in $X$
\cite{lambert2}. The modified transformation assumes the form
\begin{equation}\label{susyfermion2}
  \delta \Psi = \gamma^{\mu} \Gamma^I D_{\mu} X^I \theta - 
  \frac{1}{6} \Gamma^{I J K} \ternary{X^I, X^J, X^K} \, \theta + 
  \frac{c}{8} \Gamma^{I J K L} \Gamma^M \widetilde{G}_{I J K L} X^M \theta.
\end{equation}
Comparing with Equation \eqref{susyfermion} we note that the
modification of the transformation is in the last term. The
supersymmetry transformations for the scalars and the gauge field
remain unaltered from Equations \eqref{susyscalar} and
\eqref{susygauge}. The BPS equation  of the modified theory reads
\begin{equation}
  \delta\Psi = 0,
\end{equation}
which can be written equivalently as 
\begin{equation}\label{BPSeqn}
  \bigl[D_\mu X^I\gamma^\mu \Gamma^I - \frac{1}{6}\ternary{X^I, X^J, X^K}\G^{IJK}
  + \frac{c}{8}\Gamma^{IJKL}\G^M \widetilde{G}_{I J K L}X^M \bigr] \theta = 0.
\end{equation}
Let us note that only the anti-self-dual combination of the four-form
field appears in the last term on the left hand side, linear in $X$. 
The $R$-symmetry in this formulation is realized explicitly in
terms of the four-form field as
\begin{equation}\label{Rsymm}
R^I_J = \overline{\theta_2} \; \Gamma^{IKLM} \, \theta_1 \, 
\widetilde{G}_{KLMJ},
\end{equation}
where $\theta_1$ and $\theta_2$ are two parameters of supersymmetry
variation.


\section{Classical vacuum moduli space}\label{vac:mod}
Let us now proceed to discuss the classical vacuum
moduli space of scalars in the BLG theory without the four-form field
$G$. The action we consider, then, is the one ensuing from the
Lagrangian \eqref{action1}. The moduli space has been discussed
earlier in the literature \cite{bagger3, lambert, sen}. However, using
gauge-invariant variables to describe the moduli space as a quotient
allows us to write it as an algebraic variety over the field of real
numbers; the coordinate ring of the quotient is simply the ring of
invariant elements. Similar ideas of using gauge-invariant quantities
in studying moduli spaces of D-branes have been used earlier
\cite{fiol,subir}. The essential difference of this approach from the
earlier treatments is that in considering the part of the moduli space
arising from the flavor degrees of freedom 
we do not proceed through an intermediate
step of fixing the gauge symmetry partially.

The vacuum moduli space of scalars is the set of values of the scalars
on which the bosonic potential vanishes, modulo gauge
equivalence. Vanishing of the bosonic potential is ensured by the
vanishing of the ternary bracket,
\begin{equation} 
  \ternary{X^I,X^J,X^K} = 0.
\end{equation}
For the BLG theory satisfying Equations \eqref{strucconst} and
\eqref{symbilin}, with gauge group $\mathcal{G} = SO(4)$, this
condition reads
\begin{equation}\label{epscond}
  \epsilon^{abcd} X^I_a X^J_b X^K_c = 0,
\end{equation} 
where $I = 1, 2, \dotsc, 8$ and $a = 1, 2, 3, 4$. We shall find the
gauge-invariant moduli space of the $32$ real scalars
$X^I_a$ satisfying \eqref{epscond}.  Let us consider the ring
\[ \mathcal{R} = \R[X^I_a : I=1,2,\dotsc,8, \; a=1,2,3,4], \]
$\R$ denoting the field of real numbers 
and let us write $X = \spec \mathcal{R} = \R^8 \otimes \R^4$. 
We can construct $36$ 
gauge-invariant local variables furnishing the coordinates of the
moduli space, namely,
\begin{equation} 
y^{IJ}=\sum_{a=1}^4X^I_aX^J_a.
\end{equation}  
Let us note that due to the condition \eqref{epscond} all higher-order
invariants vanish. Denoting the ideal generated by \eqref{epscond} as
\begin{equation}
  \mathcal{J} = \langle \epsilon^{abcd} X^I_a X^J_b X^K_c \rangle \subset 
\mathcal{R}
\end{equation} 
and the subspace of $X$ defined by this ideal as $Z =
\spec(\mathcal{R} / \mathcal{J}) \hookrightarrow X$, we have the
following maps:
\begin{equation}
\label{seq}
  \R[y^{IJ}] \xrightarrow{\ f\ } \mathcal{R} \xrightarrow{\ q\ } \mathcal{R}/\mathcal{J}.
\end{equation}
Then the classical gauge-invariant moduli space written as a quotient is
\begin{equation}
\label{fullquot}
Z/\mathcal{G} = \spec\bigl( \R[y^{IJ}] / \ker( q\circ f) \bigr).
\end{equation}
Let us observe that $\ker(q \circ f) = f^{-1}(\mathcal{J})$.
Since the one-forms in $\R^4$ that generate the ideal
$\mathcal{J}$ are not invariant under the gauge group $\mathcal{G}$,
the preimage of $f$ consists of the smallest functions of the $y^{IJ}$
constructible from the one-forms, namely the metric contractions
\begin{equation}
h_{dh} \epsilon^{abcd} \epsilon^{efgh} X^I_a X^J_b X^K_c X^L_e X^M_f X^N_g,
\end{equation} 
which using \eqref{symbilin} evaluates to an ideal generated by 
$1176$ relations, \viz 
\begin{multline}
\label{IIdeal}
  \mathcal{I}_{IJKLMN} = \left\langle y^{IL}(y^{JM}y^{KN}-y^{JN}y^{KM})\right. \\
  \left. - y^{IM}(y^{JL}y^{KN}-y^{JN}y^{KL}) + y^{IN}(y^{JL}y^{KM}-y^{JM}y^{KL}) \right\rangle,
\end{multline}
where $I,J,K,L,M,N = 1, 2, \dotsc, 8$. A direct computation of
$\ker(q\circ f)$ using \emph{Macaulay~2} \cite{Macaulay2} reproduces this
ideal. The dimension of the space 
\[ \mathcal{M} = \spec \bigl( \R[y^{IJ}] \bigl/ 
\mathcal{I}_{L_1,\dotsc,L_6} \bigr)
   \text{ , \ $L_i=1,\dotsc,8$ \; for \, $i=1,\dotsc,6$.} \] 
thus obtained is found to be fifteen, using \emph{Macaulay~2}. 
The variety $\mathcal M$ is singular at the origin, which is a fixed point of
the $SO(4)$ action.

In order to relate to earlier computations \cite{lambert}
\footnote{We thank an anonymous referee for this discussion.} 
let us note that the purport of
\eqref{IIdeal} is that all the $3\times 3$ minors of the $8\times 8$
matrix $y^{IJ}$ vanish. Hence, the rank of the symmetric matrix $y^{IJ}$ is
at most two. This also makes the map $f$ in \eqref{seq} injective. 
We can, therefore, write the matrix $y^{IJ}$ as 
\begin{equation}
\label{yab}
y^{IJ} = a^Ia^J+b^Ib^J, 
\end{equation} 
where $\{a^I\}$ and $\{b^I\}$ are two linearly independent $8$-vectors,
$I=1,2,\cdots ,8$,  chosen in a suitable basis. 
This correspond to the choice of gauge employed previously
\cite{bagger3,lambert}, leading to sixteen scalars. 
However, the $SO(2)$ transformation 
\begin{equation}
\begin{split}
a^I \rt a^I\cos\theta + b^I\sin\theta,\\
b^I\rt -a^I\sin\theta + b^I\cos\theta,
\end{split}
\end{equation} 
where $\theta$ is real, keeps the matrix $y^{IJ}$ unaltered. 
Quotienting by this $SO(2)\sim U(1)$ results in the fifteen-dimensional
moduli space obtained above. 

At  this point 
let us note that so far our computations have proceeded in a manifestly 
gauge-invariant fashion. Let us also point out that 
had we considered gauging an $SO(2)$ instead of
$SO(4)$, Equation \eqref{epscond} would not have imposed any condition
on the scalars $X^I_a$. This is effectively the same as what has been
considered earlier \cite{bagger3, lambert}. In this instance too we can form
the gauge-invariant elements $y^{IJ}$ similarly as above, 
and the moduli space turns out
to be given by exactly the same equation as that for the $SO(4)$ case. In
other words, the moduli space of the $SO(2)$-theory is also
$\mathcal{M}$, with $y^{IJ}=\sum_{a=1}^2X^I_aX^J_a$. This has been verified
using \emph{Macaulay 2}.
This has indeed been the reason for
obtaining fifteen scalar moduli from the flavors upon partially fixing
the gauge group $SO(2)$. 

To reconcile with the expectations from supersymmetry, however,
the moduli space is to have one more dimension. The extra modulus arises from
the zero mode of the gauge field, as we now discuss. Since we quotient by the
group  $\mathcal{G}=SO(4)$ in \eqref{fullquot}, we may consider a fixed
non-vanishing choice of the $X$'s. In concordance with the choice \eqref{yab}
and previous results \cite{bagger3,lambert}, we may fix arbitrary real
numbers $x^I_1$ and $x^I_2$, such that
\begin{equation}
X^I=\begin{pmatrix}
x^I_1 \\x^I_2\\0\\0
\end{pmatrix},
\end{equation}
which correspond to a  point in $\mathcal M$. 
This choice of gauge and the equations of motion, in turn,
fix all the components of the gauge field, except ${\tilde{A_{\mu}}}^1_2$,
where ${\tilde{A_{\mu}}}^a_b = {\epsilon^{cda}}_b{A_{\mu}}_{cd}$. The
equation of motion for the latter now takes the form
\begin{equation}
\begin{split}
{\tilde{ F_{\mu\nu}}}^1_2 &= \pa_{\mu}{\tilde{A_{\nu}}}^1_2 -
\pa_{\nu}{\tilde{A_{\mu}}}^1_2\\
&=0,
\end{split} 
\end{equation} 
as is expected of a $U(1)$ theory. The equation of motion is solved by taking 
\begin{equation}
{\tilde{A_{\mu}}}^1_2 = \pa_{\mu}\varphi, 
\end{equation} 
where $\varphi$ is a real scalar field. 
This scalar furnishes the sixteenth dimension of
the moduli space 
\begin{equation}
{\mathcal M}_{\text{tot}}=\R\times\mathcal M. 
\end{equation} 
We thus obtain a global description of the total moduli space of the BLG
theory. 

Let us now consider the regime in which one obtains the
super-Yang-Mills theory of D2-branes from the BLG theory. Since the
considerations for the gauge fields are as above, we shall only discuss the
scalar moduli arising from the flavors. By assigning
a constant vacuum expectation value to one of the transverse scalars,
say $X^8$, and arranging the gauge field appropriately, one obtains
the Yang Mills theory with fundamental matter \cite{mukhi1}. One can
consider the vacuum moduli space of the theory of D2-branes in this
picture as well. The leading order term in the potential, proportional
to the inverse square of the Yang-Mills coupling, vanishes provided
\begin{equation}
\epsilon^{abc} X^I_b X^J_c = 0,
\end{equation}
where now $I, J = 1, 2, \dotsc, 7$. This condition is sufficient, though not 
necessary, for \eqref{epscond}, and is therefore more restrictive. 
Defining gauge-invariant coordinates as above,
 we obtain $\mathcal M$ as the seven-dimensional variety
\begin{equation}
{\mathcal M}=\spec\bigl(\R[y^{IJ}] / \mathcal{I}_{D2}\bigr),
\end{equation} 
where the ideal $\mathcal{I}_{D2}$ is given by 
\begin{equation}
  \mathcal{I}_{D2} = \left\langle (y^{IJ})^2-y^{II}y^{JJ} \right\rangle \text{ , for $I, J = 1, 2,\dotsc, 7$.}
\end{equation} 
Again, an eighth dimension is furnished by a zero mode of the gauge field, in
accordance with supersymmetry. 

\section[\texorpdfstring{$SO(1,2) \times SO(4) \times SO(4)$}{SO(1,2) x SO(1,2) x SO(1,2)}-invariant BPS equations]
        {$\bm{(SO(1,2) \times SO(4) \times SO(4))}$-invariant BPS equations}\label{so4:sol}
We shall now discuss a half-BPS configuration of the BLG theory in the
presence of the four-form field with $SO(1,2)$ world-volume symmetry
and $SO(4) \times SO(4)$ R-symmetry. We shall find that rewriting the
BPS equations using the gauge-invariant variables introduced in the
previous section facilitates the analysis of such configurations. The
$SO(4) \times SO(4)$ R-symmetry is obtained by means of a projector
$\Omega$ acting on the fermions \cite{Kim, loginov}. The projector is
obtained by seeking a realization of the irreducible spinor
representation of $\spin(8)$ in a Clifford algebra formed by monomials
of $\Gamma$-matrices. For the case at hand the projector is given by
\begin{equation}\label{bpsso4}
  \Omega = \frac{1}{4}(1+\Gamma_9 - \Gamma_{1234} - \Gamma_{5678}),
\end{equation}
where $\Gamma_{1234}$ and $\Gamma_{5678}$ denote the $\Gamma$-matrices
completely antisymmetrized in the indicated indices and $\Gamma_9 =
\Gamma_1 \dotsm \Gamma_8$.

The BPS equations can now be obtained by operating $\Omega$ on
\eqref{BPSeqn}:
\begin{equation}\label{so4operate}
  \bigl( \textstyle\frac{1}{6}\ternary{X^I, X^J, X^K} \Gamma^{IJK} - \textstyle\frac{c}{8}
  (\widetilde{G}_{I J K L}) \Gamma^{IJKL} \Gamma^{M}X^M\bigr) \, \Omega = 0.
\end{equation}
Expressing the projector explicitly in terms of the antisymmetrized 
products of the $32 \times 32$ gamma matrices of $SO(8)$, the elements 
of the BPS matrix yield the following set of equations.
\begin{equation} 
\begin{split}
\label{SO41}
\ternary{X^1,X^2,X^3} & = -\eta X^{4},\\
\ternary{X^1,X^3,X^4} & = -\eta X^{2},\\
\ternary{X^1,X^2,X^4} & = \eta X^{3},\\
\ternary{X^2,X^3,X^4} & = \eta X^{1},
\end{split}
\end{equation} 
\begin{equation} 
\begin{split}
\label{SO42}
\ternary{X^5,X^6,X^7} & = -\eta X^{8},\\
\ternary{X^5,X^7,X^8} & = -\eta X^{6},\\
\ternary{X^5,X^6,X^8} & = \eta X^{7},\\
\ternary{X^6,X^7,X^8} & = \eta X^{5},
\end{split}
\end{equation} 
where $\eta = 3 c (\widetilde{G}_{1234} - \widetilde{G}_{5678})$, in terms of the
dual four-form field,
\begin{equation} 
\begin{split}
\label{SO43}
\ternary{X^i,X^5,X^6} + \ternary{X^i,X^7,X^8} & = 0,\\
\ternary{X^i,X^5,X^7} - \ternary{X^i,X^6,X^8} & = 0,\\
\ternary{X^i,X^5,X^8} + \ternary{X^i,X^6,X^7} & = 0,
\end{split}
\end{equation} 
with $i = 1, 2, 3, 4$, and
\begin{equation} 
\begin{split}
\label{SO44}
\ternary{X^r,X^1,X^2} + \ternary{X^r,X^3,X^4} & = 0,\\
\ternary{X^r,X^1,X^3} - \ternary{X^r,X^2,X^4} & = 0,\\
\ternary{X^r,X^1,X^4} + \ternary{X^r,X^3,X^2} & = 0,
\end{split}
\end{equation} 
with $r = 5, 6, 7, 8$. 
The equations have manifest $SO(4) \times SO(4)$ R-symmetry. 

In studying the moduli space of this deformed theory, let us first
note that due to the Equations \eqref{SO41} and \eqref{SO42}, we have
to consider the order four gauge-invariants, unlike the previous case
of vanishing $\eta$. Let us denote two of the 70 order four
invariants as
\begin{equation}
  \begin{split}
    \xi_1 &= \epsilon^{abcd} X^1_a X^2_b X^3_c X^4_d, \\
    \xi_2 &= \epsilon^{abcd} X^5_a X^6_b X^7_c X^8_d.
  \end{split}
\end{equation} 
Upon multiplying both sides of \eqref{SO41} and \eqref{SO42} with a
suitable $X^I_a$ and summing over the gauge indices, we derive:
\begin{equation}
  \begin{split}\label{xi12}
    y^{11}=y^{22}=y^{33}=y^{44}=-\xi_1/\eta \\
    y^{55}=y^{66}=y^{77}=y^{88}=-\xi_2/\eta
\end{split} 
\end{equation} 
In the same fashion we also derive from these equations that
\begin{equation}
  \begin{split}\label{cross}
    y^{ij} &=0 \text{, \; for $i\neq j$ and $i, j = 1, 2, 3, 4$, and} \\
    y^{rs} &=0 \text{, \; for $r \neq s$ and $r, s = 5, 6, 7, 8$.}
\end{split}
\end{equation} 
Next, from the first equation in \eqref{SO41} we have 
\begin{equation}
  - \eta y^{45}  = \epsilon^{abcd} X^1_a X^2_b X^3_c X^5_d.
\end{equation}
Multiplying $X^3$ on both sides of the first equation in \eqref{SO44}
with $r=5$ then yields, upon summing over the gauge indices,
\begin{equation}
\label{411}
  \epsilon^{abcd} X^1_a X^2_b X^3_c X^5_d = -\epsilon^{abcd}X^5_aX^3_bX^4_cX^3_d = 0.
\end{equation} 
Hence we conclude that $y^{45} = 0$. Proceeding similarly with the
other equations, we obtain
\begin{equation}\label{yri}
  y^{ri} = 0 \text{ \; for all $i=1,2,3,4$ and $r=5,6,7,8$.}
\end{equation} 
This, in turn, makes all the order four invariants apart from $\xi_1$ and
$\xi_2$ vanish. 
Furthermore, from the first equation in \eqref{SO41} we get, upon squaring,
\begin{equation}\label{y45}
  \epsilon^{abcd} \epsilon^{pqr}_d X^1_a X^2_b X^3_c X^1_p X^2_q X^3_r = \eta^2 X^4_dX^4_d,
\end{equation} 
where repeated indices are summed. This leads to 
\begin{equation}
  y^{11} y^{22} y^{33} = \eta^2 y^{44},
\end{equation} 
which, using \eqref{xi12}, yields
\begin{equation}\label{xi1eq}
  \xi_1(\xi_1^2-\eta^4) = 0.
\end{equation} 
Similarly, from \eqref{SO42} and \eqref{xi12} we derive
\begin{equation}
\label{xi2eq} 
\xi_2(\xi_2^2-\eta^4)=0. 
\end{equation} 
Finally, by squaring $\epsilon^{abcd}X^1_aX^2_bx^3_cX^5_d$ and using
\eqref{411}, we obtain
\begin{equation}
  y^{11} y^{22} y^{33} y^{55} = 0,
\end{equation}
that is,
\begin{equation}\label{xixi}
  \xi_1^3 \xi_2 = 0.
\end{equation} 
This equation, along with \eqref{xi1eq} and \eqref{xi2eq}, implies
that one of the $\xi$'s have to vanish. If we choose $\xi_2 = 0$,
then using \eqref{xi12} and recalling that $y^{II}$, for each $I$,
is a  sum of squares of real numbers, we conclude that
\begin{equation}
  X^5 = X^6 = X^7 = X^8 = 0.
\end{equation} 
From \eqref{xi1eq} we then have $\xi_1 = \pm\eta^2$ and by \eqref{xi12} we are thus left with
\begin{equation} 
  y^{11} = y^{22} = y^{33} = y^{44} = \pm\eta,
\end{equation} 
with the sign chosen such that these invariants are positive-definite
for a non-vanishing $\eta$. All other gauge invariants vanish. We
conclude that the gauge-invariant moduli space is zero-dimensional.
An explicit BPS configuration is then furnished by, for example,
choosing the four scalars $X^i$, $i = 1, 2, 3, 4$ to be proportional
to the four basis vectors in the Euclidean space $\R^4$ of the
gauge indices.

\section{Conclusion}\label{concl}
We have examined the classical vacuum moduli space of
scalars of the BLG theory. By considering gauge-invariant combinations of
scalars we obtain a global description of the
sixteen-dimensional moduli space as the product of a fifteen-dimensional 
real singular
algebraic variety $\mathcal M$ and the real line, the
latter arising from a zero-mode of the three-dimensional gauge field. 
We then use the gauge-invariant
variables to study a half-BPS configuration of the BLG theory deformed
by a four-form background field. The configuration has
$SO(1,2)$-world-volume symmetry and $SO(4)\times SO(4)$ R-symmetry.
In this case, however, we have to consider invariants higher than
order two in the scalars $X$. We obtain an explicit solution derived
from the fact that $\mathcal M$ is 
zero-dimensional, parametrized by the four-form field.  We find that
the BPS equations allow only four of the eight scalars, corresponding
to one of the $SO(4)$ R-symmetry factors, to be non-zero.

This method of forming gauge invariants is extremely useful in
studying moduli spaces of gauge theories. They serve also as a useful
tool for studying BPS configurations by constructing explicit
solutions. It seems that configurations with various other
R-symmetries can also be studied in this fashion. We hope to report on
these issues in future.


\section*{Acknowledgments}
\addcontentsline{toc}{section}{Acknowledgments}
TK and KR are partially supported by Grant 2008/R2 of the Royal
Society. They would like to thank the PI's Elizabeth Gasparim and Pushan
Majumdar for being generous with the grant 
under which they enjoy a fruitful exchange between the IACS,
Calcutta and the University of Edinburgh, UK. 
SPC is supported by the Department of Science and Technology, Government of
India. 
SC wishes to thank the Department of Physics, IACS, for warm hospitality
during a major part of this work.
We thank Anirban Basu, Bobby
Ezhuthachan, Jos\'{e} Figueroa-O'Farrill, Elizabeth Gasparim, Sachin Jain,
Sudipta Mukherji for several helpful discussions and the anonymous referees
for important suggestions.


\end{document}